\documentclass[conference]{IEEEtran}
\IEEEoverridecommandlockouts

\usepackage{cite}
\usepackage{amsmath,amssymb,amsfonts}
\usepackage{algorithmic}
\usepackage{graphicx}
\usepackage{textcomp}
\usepackage{xcolor}
\def\BibTeX{{\rm B\kern-.05em{\sc i\kern-.025em b}\kern-.08em
    T\kern-.1667em\lower.7ex\hbox{E}\kern-.125emX}}

\usepackage{tabularray}
\usepackage{orcidlink}
\usepackage{csquotes}
\usepackage{newfloat}

\DeclareFloatingEnvironment[fileext=lst,placement={!htbp},name=Listing]{listing}

\DeclareQuoteStyle{italics}
{\itshape\textquotedblleft}  %
  {\itshape\textquotedblright} %
  {\normalshape\textquoteleft}     %
  {\textquoteright}  

\usepackage{glossaries}
\glsdisablehyper
\setacronymstyle{long-short}
\newacronym{llm}{LLM}{large language model}
\newacronym{saas}{SaaS}{software as a service}
\newacronym{ddd}{DDD}{domain-driven design}
\newacronym{api}{API}{application programming interface}

\usepackage{listings}

\lstdefinestyle{prompt}{
    basicstyle=\ttfamily\scriptsize,
    frame=single,
    framerule=0.5pt,
    breaklines=true,                %
    breakatwhitespace=false,
    columns=fullflexible,           %
    keepspaces=true,                %
    showstringspaces=false,
    xleftmargin=0.5em,              %
    xrightmargin=0.5em,
}

\lstdefinestyle{inlineprompt}{
    style=prompt,      %
    basicstyle=\ttfamily\normalsize,
}

\usetikzlibrary{positioning,calc}

\newcommand{\generated}[1]{}
    
\begin{document}
\bstctlcite{BSTcontrol}
\setquotestyle{italics}

\title{Automating Domain-Driven~Design: Experience~with~a~Prompting~Framework\\
}

\author{\IEEEauthorblockN{1\textsuperscript{st} Tobias Eisenreich\,\orcidlink{0009-0004-7168-251X}}
\IEEEauthorblockA{\textit{Chair of Software Engineering} \\
\textit{Technical University of Munich}\\
Heilbronn, Germany \\
tobias.eisenreich@tum.de}
\and
\IEEEauthorblockN{2\textsuperscript{nd} Husein Jusic}
\IEEEauthorblockA{\textit{Engineering Team} \\
\textit{FTAPI Software GmbH}\\
Munich, Germany \\
h.jusic@ftapi.com}
\and
\IEEEauthorblockN{3\textsuperscript{rd} Stefan Wagner\,\orcidlink{0000-0002-5256-8429}}
\IEEEauthorblockA{\textit{Chair of Software Engineering} \\
\textit{Technical University of Munich}\\
Heilbronn, Germany \\
stefan.wagner@tum.de}
}

\maketitle
\IEEEpeerreviewmaketitle

\begin{abstract}
\Gls{ddd} is a powerful design technique for architecting complex software systems. This paper introduces a prompting framework that automates core \gls{ddd} activities through structured \gls{llm} interactions. We decompose \gls{ddd} into five sequential steps: (1) establishing an ubiquitous language, (2) simulating event storming, (3) identifying bounded contexts, (4) designing aggregates, and (5) mapping to technical architecture. In a case study, we validated the prompting framework against real-world requirements from FTAPI’s enterprise platform. While the first steps consistently generate valuable and usable artifacts, later steps show how minor errors or inaccuracies can propagate and accumulate. Overall, the framework excels as a collaborative sparring partner for building actionable documentation, such as glossaries and context maps, but not for full automation.
This allows the experts to concentrate their discussion on the critical trade-offs. In our evaluation, Steps 1 to 3 worked well, but the accumulated errors rendered the artifacts generated from Steps 4 and 5 impractical.
Our findings show that \glspl{llm} can enhance, but not replace, architectural expertise, offering a practical tool to reduce the effort and overhead of \gls{ddd} while preserving human-centric decision-making.
\end{abstract}


\glsresetall %

\section{Introduction\generated{ (1 page)}}

Designing software architecture is inherently challenging. It needs to cater to complex business rules, support quality attributes, and integrate technical solutions. \Gls{ddd} and associated techniques promised to lighten this load, and they do; \gls{ddd} is now one of the most used and most effective approaches to design complex business applications~\cite{ozkan2025domain}. However, this comes at a cost: designing an architecture for a complex software system requires many stakeholders to collaborate in time-intensive workshops to align everything.

With \glspl{llm} now demonstrating remarkable reasoning capabilities, they could assist with tedious tasks in the \gls{ddd} process. \Gls{llm}-generated artifacts could be used in these workshops to guide stakeholders toward the most challenging parts of the process, avoiding time-consuming work on simple aspects. This could potentially both shorten workshop times and increase the quality of the designed architecture. The idea to automate parts of the \gls{ddd} process leads to our research question:

\begin{quote}
    \textbf{RQ:} How effectively can LLMs automate different stages of the Domain-Driven Design process, and where does human expertise remain essential?
\end{quote}

We developed a prompt framework to investigate the capabilities of \glspl{llm} to assist with the \gls{ddd} process.
While the reasoning capabilities of \glspl{llm} are remarkable, they still show detrimental behaviors like hallucinations, and can sometimes struggle with very complex tasks. Therefore, this prompting framework fosters collaboration between a software architect and the \gls{llm}, refining the design collaboratively and ideally resolving hallucinations immediately.

To evaluate the prompting framework, we collaborated with FTAPI Software GmbH. 
They are currently modularizing their decade-old enterprise platform. Over time, parts of the platform degenerated into a \textit{big ball of mud}, making the maintenance and continued development challenging.
FTAPI has already refactored a component called SecuRooms using \gls{ddd} without the framework we contribute in this paper. 
Applying our prompt framework, we generated \gls{ddd} artifacts for SecuRooms. We compared them against the previously modularized SecuRooms architecture. For the upcoming modularization of the next component, SecuMails, we generated artifacts to support the architects in the \gls{ddd} process.

Our evaluation shows that \glspl{llm} are very effective in establishing a ubiquitous language and relating terms to each other in a simulated event storming. These capabilities alone can substantially accelerate the \gls{ddd} process. Furthermore, while the \glspl{llm} can struggle with identifying sensible bounded contexts or aggregates, we found that they can bring new perspectives to the table that the human expert software architects had not considered before.

By adopting our five-step prompting framework, software architects can accelerate their \gls{ddd} process while introducing new ideas they might not have considered. An important prerequisite is the availability of the system requirements in a textual format.

The rest of this paper is structured as follows:
Section \ref{sec:background} explains the problem context, the foundational research, and related work. Section \ref{sec:automating} describes our proposed five-step framework and explains how to implement it. Section \ref{sec:methodology} explains the methods that we used to evaluate the framework in a case study. Section \ref{sec:results} details the results of our evaluation, and Section \ref{sec:discussion} discusses them. With Section~\ref{sec:conclusion}, we conclude the paper.

\section{Background and Related Work\generated{ (1 page)}}\label{sec:background}

Over the last 20 years, \gls{ddd}~\cite{evans2004domain, vernon2016domain} has evolved into one of the popular methods for software architecture design and is an effective approach for developing high-quality software architectures~\cite{ozkan2025domain}. It places the main focus of the architecture on the application domain and the business rules, leaving technical details such as programming language, libraries, and frameworks open as implementation details. This leads to architectures that are comparatively easy to maintain, especially in systems with complex requirements and business rules~\cite{ozkan2025domain}. For our case study with a complex and interleaved business domain, this approach fits very well.

Very recently, the \glspl{llm} have shown major advancements. These new possibilities spread to many research fields and are readily used by practitioners.
Using \glspl{llm}, it is important to formulate the prompts concisely, so that they tell the \gls{llm} exactly what output is expected. This field of prompt engineering explored many possibilities of doing so~\cite{schulhoff2025prompt}. One consensus is the use of specialized prompt templates for recurring problems. With these, the expert writing the prompts does not need to concern themselves with optimizing it, as others have already done this. Our contribution is a collection of such prompt templates for use in the \gls{ddd} process.

Many researchers and practitioners have applied \glspl{llm} to software architecture. Esposito et al.~\cite{esposito2026generative}, and Bucaioni et al.~\cite{bucaioni2025artificial} conducted literature reviews providing a comprehensive overview of the topic. Overall, the related work shows general challenges and opportunities of \gls{llm}-based approaches~\cite{dirocco2024uselargelanguagemodels}, confirms that \gls{llm}-based approaches outperform rule-based approaches~\cite{tagliaferro2025leveraging} and are continuously improving~\cite{yang2024multistep,chaaben2025utility}, but currently lack the accuracy of human architects and are impractical for full automation~\cite{cervantes2025llmassistedapproachdesigningsoftware,chen2023automated}.
 
Dhar et al.~\cite{dhar2025draftingarchitecturaldesigndecisions} introduce \textit{Domain-specific Retrieval Augmented Few-shot Tuning} (DRAFT) to generate architectural decision records (ADRs), leveraging both fine-tuning and retrieval-augmented generation. They show that DRAFT outperforms other approaches in effectiveness and efficiency on a large dataset of ADRs from GitHub.

Helmi~\cite{helmi2025arlotailorableapproachtransforming} introduces ARLO, a \gls{llm}-based tool that analyzes natural-language requirements and recommends architectural patterns from a database based on their known characteristics. It matches these known characteristics with the quality attributes derived from the requirements, choosing the optimal architecture using linear programming.

Tagliaferro et al.~\cite{tagliaferro2025leveraging} derive UML component diagrams from informal specifications. Using a quantitative metric against a developed ground truth, they found that \gls{llm}-based approaches outperform traditional rule-based approaches but still lack the accuracy needed for deployment in real-world scenarios.

Cervantes et al.~\cite{cervantes2025llmassistedapproachdesigningsoftware} employ \gls{llm}-assistants to synthesize software architecture using attribute-driven design (ADD) in collaboration with a human architect. Similar to our work, they find that human oversight and iterative refinement remain critical for developing architectures well aligned with established solutions.

To the best of our knowledge, there is no research on enhancing \gls{ddd} as a process by utilizing \glspl{llm}. However, there are commercial solutions for practitioners exploring the area: Qlerify\footnote{\url{https://www.qlerify.com/domain-driven-design-tool}} offers \gls{llm} support for practitioners in their \gls{ddd} processes. Acknowledging the real-world importance of this research field, we reaffirm our ambition to improve the scientific understanding of automated or assisted \gls{ddd}. In this work, we show how using \glspl{llm} for \gls{ddd} is feasible. At the same time, we show areas for further improvement in the process.

\generated{
    Do:
        First paragraph: Your specific architectural challenge (monolith modernization)
        Second paragraph: Why existing approaches fail (cite ONLY 2-3 most relevant papers)
        Third paragraph: What practitioners need (not what academics study)
    Critical: Add: "For readers facing similar challenges: This approach works best when [criteria]"
    Why it works: Satisfies CFP's "Problem Context and Industry Impact" criterion while staying practitioner-focused
}

\section{Automating Domain-Driven Design\generated{ (2 pages) [Replaces Methodology]}}\label{sec:automating}

\begin{figure}[tbp]
    \centering
    \begin{tikzpicture}[
            node distance=0.6cm,
            phase/.style={
                rectangle, rounded corners=4pt, minimum width=7.2cm,
                minimum height=1cm, text centered, draw=black, font=\small\bfseries, text width=7cm
            },
            arrow/.style={thick,->,>=stealth, color=blue!70},
        ]

        \node (requirements) [phase, fill=gray!10] at (0.0, 0.0) {Requirements Input};
        
        \node (phase1) [phase, below=of requirements, fill=blue!10] {
            Step 1: Ubiquitous Language Establishment\\
            \scriptsize Extract domain vocabulary, define glossary
        };
        
        \node (phase2) [phase, below=of phase1, fill=red!10] {
            Step 2: Event Storming Simulation\\
            \scriptsize Identify events, commands, actors
        };
        
        \node (phase3) [phase, below=of phase2, fill=orange!10] {
            Step 3: Bounded Context Identification\\
            \scriptsize Group concepts into cohesive contexts
        };
        
        \node (phase4) [phase, below=of phase3, fill=purple!10] {
            Step 4: Aggregate Design\\
            \scriptsize Define aggregates, entities, invariants
        };
        
        \node (phase5) [phase, below=of phase4, fill=green!10] {
            Step 5: Technical Architecture Mapping\\
            \scriptsize Design ports, adapters, infrastructure
        };
        
        \node (output) [phase, below=of phase5, fill=gray!10, font=\bfseries] {
            Complete Domain Model\\
            \scriptsize Bounded Contexts, Aggregates, Architecture
        };

        \path let
            \p1 = (requirements.north),
            \p2 = (output.south),
            \n{h} = {\y1 - \y2}
        in
        node (systemprompt) [phase, rotate=90, minimum width=\n{h}, yshift=4.7cm, fill=yellow!10] at ($(requirements)!0.5!(output)$) {
            System Prompt\\
            \scriptsize Define persona, guide through questions, intervene on red flags
        };
        
        \draw[arrow] (requirements.south) -- (phase1.north);
        \draw [arrow] (phase1.south) -- (phase2.north);
        \draw [arrow] (phase2.south) -- (phase3.north);
        \draw [arrow] (phase3.south) -- (phase4.north);
        \draw [arrow] (phase4.south) -- (phase5.north);
        \draw [arrow] (phase5.south) -- (output.north);
    
    \end{tikzpicture}
    \caption{Five-Step LLM-Assisted DDD Analysis Workflow.}
    \label{fig:refined-ddd-workflow}
\end{figure}
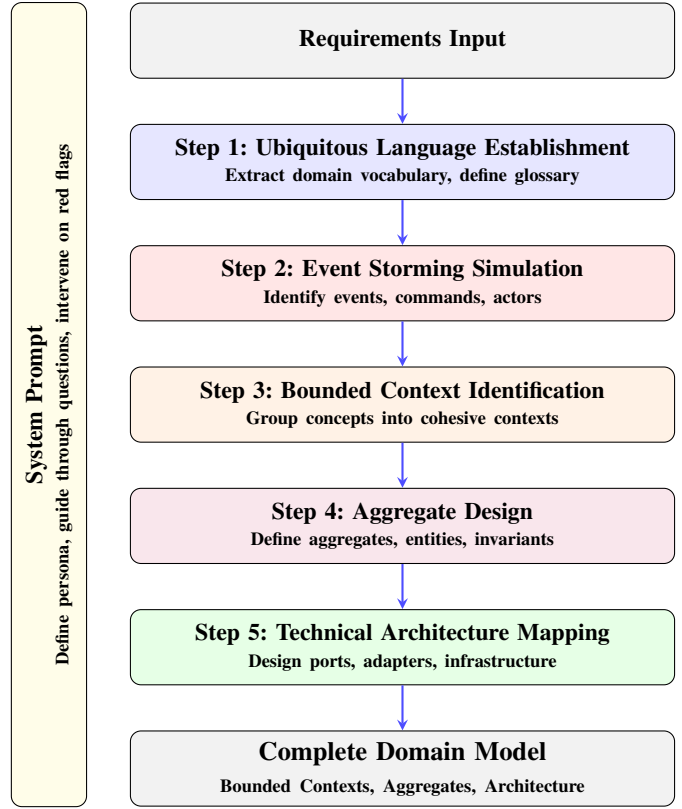

To automate \gls{ddd}, we break the process down into five steps (see \autoref{fig:refined-ddd-workflow}). They are designed to follow the natural progression of the \gls{ddd} practice. Each step is implemented using a single prompt. The last step yields a complete application design:
Step 1 establishes an ubiquitous language. Step 2 simulates an event storming. Step 3 identifies the bounded contexts. Step 4 defines the aggregates, entities, and invariants. Finally, Step 5 maps it all to a technical architecture.
These five steps were separated based on our intuition and the experts' experience at FTAPI. The prompts were crafted carefully, following prompt engineering guidelines~\cite{schulhoff2025prompt}.%

The prompt for each step is executed in the same chat, using a custom system prompt. The requirements are attached to the chat as a text file. When the \gls{llm} asked questions in response to the prompt, they were answered accordingly before proceeding with the prompt of the next step. The \gls{llm} was additionally prompted to visualize the output using PlantUML. Since the full prompts are too long to include, they are available in the supplementary data~\cite{supplementarymaterial}. The prompt for Step~2 is included as an example in \autoref{lst:prm:event-storming}.

\subsection{System Prompt}

The system prompt establishes the \gls{llm} as a \textit{Senior \gls{ddd} Specialist} and \textit{Architectural Sparring Partner}, using an architect persona~\cite{maranhao2024prompt}. This detailed persona specification provides contextual grounding for the expected level of architectural reasoning and encourages critical thinking and challenging assumptions.

The system prompt guides the \gls{llm}'s analytical process: it should enforce \gls{ddd} best practices, engage in collaborative modeling, and actively challenge ambiguous concepts. It lists red flags that should trigger an intervention, and encourages understanding the requirements by asking questions. During our prompt design, we found that without using such a system prompt, the \gls{llm} was prone to losing focus.

\subsection{Step 1: Establishing the Ubiquitous Language}

In the first step, the \gls{llm} systematically extracts and defines the core domain vocabulary from the requirement specification. This guides all subsequent analysis to operate within a consistent linguistic framework.

The prompt instructs the model to identify the key terms. It is asked to generate a business-focused definition for each term, identify its context, link related terms, and ask clarifying questions if the requirements document leaves room for ambiguity or uncertainty.
The architect using the prompt framework can subsequently clarify these uncertainties to improve the glossary. Once the architect has the impression that these are sufficiently clarified, they can advance to Step~2.

\subsection{Step 2: Event Storming Simulation}

\begin{listing}[tbp]   %
\centering
\begin{lstlisting}[
    style=prompt,
]
Based on the ubiquitous language we've established, let's conduct an Event Storming session:

1. Identify all Domain Events (things that happen) in chronological order
2. For each event, identify:
    - The Command that triggers it
    - The Actor/Role who initiates the command
    - Any Policies/Rules that apply
    - The Aggregate that handles it
3. Look for temporal boundaries and parallel processes
4. Create a visual flow showing the event stream

Format as:
Actor -> Command -> Aggregate -> Event(s) -> Policy/Reaction -> Next Command

Highlight any areas where the flow seems unclear or where multiple interpretations exist.
\end{lstlisting}
\caption{Prompt for the event storming simulation.}
\label{lst:prm:event-storming}
\end{listing}

\textit{Event Storming} was developed in the context of \gls{ddd}, and popularized by Vernon~\cite{vernon2016domain}. Typically, an Event Storming is conducted to identify the ubiquitous language and discover the relationships between the key terms. In this automated workflow, and in contrast to Event Storming as originally intended, the establishment of the ubiquitous language is separated and occurs in Step~1 to reduce the complexity of each step.

The prompt in \autoref{lst:prm:event-storming} guides the \gls{llm} to systematically identify domain events in chronological order, relating each event to its triggering command(s), responsible actor(s), applicable policies, and handling aggregates. This phase transforms the static vocabulary of the ubiquitous language into a dynamic understanding. Some parts, such as aggregates, may not be accurate at this step, but are revisited later. Event storming is primarily intended to identify and structure all aspects of the domain related to the requirements.

\subsection{Step 3: Identifying the Bounded Contexts}

After the event storming, the relations between the key terms are clear. Usually, some of the terms are clustered together. These clusters are now formalized as the domain's bounded contexts.

The prompt guides the \gls{llm} to identify these bounded contexts and detail them, defining each context’s core purpose and key aggregates. All terms from the ubiquitous language established in Step~1 are grouped into these contexts. The bounded contexts identified establish the high-level architectural boundaries.

\subsection{Step 4: Designing the Aggregates}

Each bounded context comprises one or multiple aggregates. These define invariants and business rules that apply to the elements they contain. While the \gls{llm} has already identified aggregates in Step 2, we revisit them here to provide a clear definition for each.

To refine the aggregates, the prompt instructs the \gls{llm} to identify the aggregate roots. These are entities that control access to the aggregate and enforce business rules.
The prompt also instructs the model to identify the business invariants, the events issued, and the actions associated with the aggregate.
Specifically, the \gls{llm} should challenge aggregates that do not protect specific business invariants or are not small enough, and refine them as needed.

\subsection{Step 5: Technical Architecture Mapping}

Finally, the technical aspects of the architecture need to be designed.
The prompt asks the \gls{llm} to generate the necessary technical components, like anti-corruption layers and \glspl{api}. Furthermore, the model should explain how each technical decision supports the domain model.
We found that the \glspl{llm} struggled with this step. Providing additional guidance on architectural patterns in the prompt improved generation. The FTAPI experts recommended mentioning hexagonal architectures, as well as repository and specification patterns, in the prompt to guide the \gls{llm}.

\subsection{How to Implement this Prompting Framework}

Based on our evaluation, we recommend the following workflow when using the prompting framework. Read at least the \nameref{sec:results} and \nameref{sec:discussion} Sections in addition to avoid pitfalls, please.
\begin{enumerate}
    \item Retrieve the latest prompt template versions from the supplementary material~\cite{supplementarymaterial}.
    \item Collect detailed requirements into a \gls{llm}-readable file format, for example Markdown.
    \item Insert the system prompt into your \gls{llm}-tool.
    \item Insert the prompt for Step~1; attach the requirements file.
    \item Answer the \gls{llm}'s questions until the output is accurate.
    \item Proceed with the consecutive steps in the same manner.
\end{enumerate}

\section{Methodology}\label{sec:methodology}

To assess the effectiveness of our framework and identify where human expertise remains essential, we apply our framework in an industrial case study.

\subsection{FTAPI Software GmbH}

FTAPI Software GmbH is developing the SecuTransfer platform, a software specializing in secure file transfer methods. The platform consists of four integrated products handling emails, forms, cloud storage, and automated workflows. It is historically implemented in a monolithic architecture.

FTAPI is currently modularizing this monolithic platform. They have already modularized SecuRooms, the core component handling cloud storage, using \gls{ddd} without \gls{llm} support.
This constellation allows the experts to compare the modularized design of SecuRooms with the design suggested by the \gls{llm} using the prompting framework. 
A second core component for handling emails, SecuMails, is an organically grown monolith and the next component to be modularized.
Using the prompting framework, we generated a potential design for SecuMails and asked the experts to analyze it and identify potential issues.

FTAPI provides a redacted version of the requirements for both SecuMails and SecuRooms. They are available in the supplementary data~\cite{supplementarymaterial}. These requirements are representative of the internal specifications; some details may differ from the actual production system.

\subsection{Data Collection}

One expert executed the five-step workflow on three different \glspl{llm}. We selected \textit{Claude Opus 4.1} by Anthropic, \textit{Gemini 2.5 Pro} by Google, and \textit{GPT 5.0} by OpenAI. These models were readily available in the company context and, in our experience, generally perform well. The expert engaged in a dialog with the \glspl{llm} and answered all questions asked, following the prompts. For each \gls{llm} and each workflow step, the final \gls{llm} output was collected for analysis in the interviews.

\subsection{Interviewing the Experts}

To thoroughly evaluate the generated data, we interviewed three architecture experts at FTAPI who are familiar with both SecuRooms and SecuMails. One of them is a technical leader, the other two are backend developers of the examined products. The interviews were based on the \gls{llm} outputs of each step. Each interview lasted about one hour. The full interview structure, including guiding questions, is available in the supplementary material~\cite{supplementarymaterial}. The interviews were recorded and subsequently transcribed using HappyScribe\footnote{\url{https://www.happyscribe.com}}.

\subsection{Data Analysis}

Using thematic analysis, the interview transcripts were clustered into five topics: One topic for each of the five steps of the prompting framework, capturing the impressions of the experts. The findings are reported in Section~\ref{sec:results}.

\section{Results\generated{ (2 pages)}}\label{sec:results}

This section presents our findings from the interviews, organized by the five workflow steps. The complete set of diagrams generated and used for the evaluation is available in the supplementary materials~\cite{supplementarymaterial}; a subset is included in the paper for illustration.

\subsection{Findings from Step 1: Ubiquitous Language}
\begin{table*}[tbp]
\caption{Three terms extracted into an ubiquitous language by ChatGPT for SecuRooms.}
\label{tab:step-1}
\begin{tblr}{
    colspec={X[0.6]X[3]X[2]X[1.5]X[3]},
    row{1}={font=\bfseries},
    row{2,Z}={halign=l},
    hline{1,2,Z},
}
Term & Definition & Business Context & Related Terms & Questions / Clarifications Needed \\
File & A single digital document or object (e.g., PDF, image, archive) stored in a Data Room. & Uploaded, downloaded, versioned, viewed or annotated. & File Version, Upload, Download, PDF-Viewer & Are there file-type restrictions per customer? Are files immutable after upload unless versioned? \\
File Version & A saved snapshot of a File at a point in time; previous versions can be restored. & Maintained automatically when a File is changed. & File, Version History, Restore & How many versions are retained? Retention policy per Data Room? \\
Owner & A role with the highest authority over a Data Room: can delete the Data Room, manage access and view audit info. & Assigned per Data Room; visible in Admin Console. & Roles, Permissions, Admin & Can there be multiple Owners? What actions are reserved exclusively to owners vs. administrators?
\end{tblr}
\end{table*}
In general, Step 1 worked well: The main goal, the extraction of the ubiquitous language as a basis for further discussion and analysis, was achieved by all three \glspl{llm} tested. \autoref{tab:step-1} shows an example for three terms generated by ChatGPT. The \gls{llm} asked follow-up questions about the terms, aiming to sharpen and clarify their definitions. The architect could then answer these questions to refine the terms as needed. The experts found this step very valuable, as it clearly defines the terms used later in the process, avoiding ambiguities.

However, the experts also noted that \enquote{some [terms] were too generic and would need [further] refinement}, and some \enquote{feel a bit artificial compared to how we usually talk}.

\subsection{Findings from Step 2: Event Storming}
\begin{figure}[tbp]
    \centering
    \includegraphics[width=\linewidth]{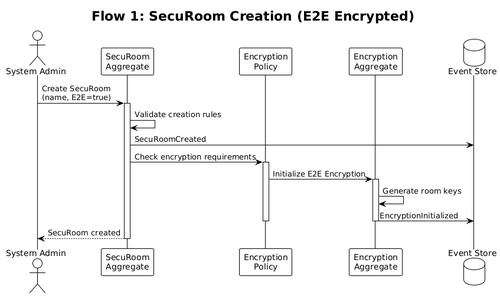}
    \caption{Event Storming: PlantUML diagram for the creation of a SecuRoom by Claude.}
    \label{fig:step-2}
\end{figure}
The evaluations revealed a mixed picture for the Event Storming step. On the positive side, experts agreed that the \gls{llm} captured the core business events. \autoref{fig:step-2} shows a PlantUML visualization of a business flow found by Claude. However, the experts also pointed out gaps, especially in the coverage of edge cases and technical events that arise in everyday implementation. As one expert explained, some of the missing details were events that \enquote{we deal with constantly in operations, but which may not show up in the main business description.} This stresses the importance of complete and accurate requirements when using this automated workflow to avoid missing business flows.

Importantly, the interactive setup of the event storming exercise surfaced these gaps: \enquote{The \gls{llm} is asking very good questions going through the steps.} By engaging in this questioning, the process uncovered events that might otherwise have been missed in a static or purely automated analysis of new features. 

Overall, Step 2 demonstrated that while \glspl{llm} can provide a solid baseline of domain events, human expertise remains important for capturing the whole picture. The structured nature of the simulation gave confidence in the correctness of the main workflows, while the expert review ensured that overlooked or implicit events were also considered.

\subsection{Findings from Step 3: Bounded Contexts}
\begin{figure*}[tbp]
    \centering
    \includegraphics[width=\linewidth]{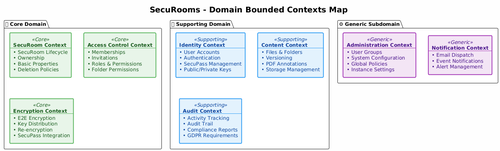}
    \caption{Bounded contexts for SecuRooms as proposed by Claude.}
    \label{fig:step-3}
\end{figure*}
The identification of the bounded contexts was partially effective, but it also revealed limitations.
The \gls{llm} was effective at identifying theoretically valid boundaries but sometimes failed to consider the practical dependencies and tight coupling that shape real-world architectures.

Compared to the existing modularization of SecuRooms, the experts found that the \gls{llm} sometimes suggests splitting into more fine-grained bounded contexts than the experts deemed practical. For example, Claude suggested splitting the \textit{Access Control} from the \textit{Identity} context (see \autoref{fig:step-3}). An expert noted: \enquote{I'm just thinking about what would be the point of dividing it up\dots{} it definitely depends on each other.} In our evaluation, the \glspl{llm} tend to optimize for theoretical clarity and separation of concerns, while human architects also consider practical factors such as coupling and operational overhead. This might be caused by the prompting, which encourages the split into small bounded contexts: \enquote{[\dots] \lstinline[style=inlineprompt]{Question any contexts that seem too large or have unclear boundaries.}}

On other occasions, the experts agreed with the smaller split which the \gls{llm} proposed: \enquote{I like the idea of the encryption context being branched out into its own context. It is somehow already a domain, but not yet worked out enough.} Another expert added that the encryption in the actual system \enquote{is not really set up as a domain. [\dots] But I wouldn’t [set it up like] that from the start.}

All in all, Step 3 seems to be a good fit for a sparring partner approach, supporting experienced architects and accelerating the design process.

\subsection{Findings from Step 4: Aggregates}
\begin{figure*}[tbp]
    \centering
    \includegraphics[width=\linewidth]{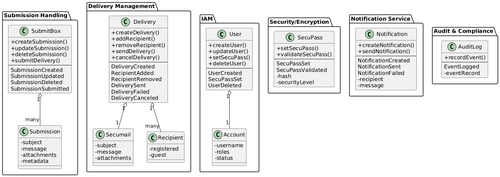}
    \caption{SecuMails aggregates generated by ChatGPT.}
    \label{fig:step-4}
\end{figure*}
The \gls{llm}'s proposed aggregates did not align well with the real-world aggregates from SecuRooms. This is partly induced by the misalignment of the results from Step 3, which serve as input to all subsequent steps. At the same time, this makes comparison-based analysis impossible.
Also, the level of detail varied significantly between the \glspl{llm}, with Claude 4.1 being the most detailed. The expert found that at least some of the generated aggregates aligned with the existing system.

\autoref{fig:step-4} shows the aggregates suggested by ChatGPT for the SecuMails requirements set. The experts found them to be sensible in general. However, they also noted that some aggregates lacked the contextual detail and depth necessary to accurately reflect the core domain.

\subsection{Findings from Step 5: Technical Architecture Mapping}\label{ss:results-step-5}
The final architecture mapping showed little usefulness. Since the alignment with the actual system had mostly been lost by this point, the experts found it difficult to judge the mapping of the technical architecture. Most of the generated outputs consisted of very shallow explanations with little substantive detail. An attempt to generate a PlantUML diagram of the architecture resulted in a rather cluttered image with neither a lot of detail nor a good visual impression. It is available in the supplementary material~\cite{supplementarymaterial} for reference. Improving the prompting helped to a certain degree -- before, the results were worse --, but did not bring a breakthrough. Further research is needed.

\section{Discussion}\label{sec:discussion}

This section discusses the results and sets them into context. Additionally, we discuss implications for practice and future research, as well as limitations to the generalizability of our findings.

\subsection{Interpretation of the Results}
It is well known that \glspl{llm} can assist very well with linguistic tasks, and thus it is not surprising to see an excellent glossary, which is little more than a list of the terms appearing in the requirements. It is worth noting, though, that the \glspl{llm} also excelled at asking critical questions about the project requirements and terms, clarifying the initial requirements. Although this was not evaluated scientifically, the authors noticed during the development of the prompting framework that the \gls{llm} demonstrated a strong understanding of the project domain, providing accurate guidance on joining or demarkating terms in the glossary.

The same is true for the second step: Relating terms to each other is a key competence of \glspl{llm}, so it is not surprising that they excel at the task. Since Event Storming is usually a collaborative effort, it is important to assess the \gls{llm}'s ability to uncover all domain concepts that may have been missed in Step~1. Our experts suggested that this might not be the case, voicing concerns that some business flows had not been identified. For this step specifically, it might be beneficial to employ multiple orchestrated \glspl{llm} with different personas, simulating the stakeholders discussing with each other. This could be tried and evaluated in future work.

In Step~3, the experts had their first substantial disagreements with the generated artifacts. They mostly boil down to the granularity of the extracted contexts: All \glspl{llm} suggested more granular bounded contexts than those actively used at FTAPI. However, the FTAPI experts also noted that, in some instances, their architecture would benefit from the additional bounded contexts generated. In other instances, the experts agreed that more granular contexts would not make sense for their current product. This suggests that the generated bounded contexts might be beneficial for stakeholder discussions.
At the same time, we assume that the prompt for Step~3 could be improved. Regarding granularity, the prompt instructs the \gls{llm} like this:

\noindent\begin{minipage}{\columnwidth}
\begin{lstlisting}[style=prompt]
Question any contexts that seem too large or have unclear boundaries.
\end{lstlisting}
\end{minipage}

This guides the \gls{llm} to break up contexts into smaller units, but never to join small contexts together if the splits become too granular. A refined prompt could improve the bounded contexts in a future version of the prompt framework.

The results show substantial limitations on the assistance in Step~4. We want to discuss two potential reasons:
First, we got the impression that even minor inaccuracies accumulated over multiple steps. We presume that these accumulated errors reach their tipping point around Step~4. The lack of iterative refinement in our case study may lead to such errors. This is a shortcoming we aim to address in future work.
Second, it is a known limitation of \glspl{llm} that they have a limited context window. A large requirements document, followed by the development of a glossary, an Event Storming protocol, and bounded contexts, along with interactive refinements of each step, might exceed this context window, especially since it has been shown that \glspl{llm} have an attention bias towards more recent tokens and can lose information before they drop out of the context window. This could be improved in future work by using a fresh context for each step and supplying only the necessary data in that context. For Step~4, it would be sufficient to include the requirements documents, the glossary, the event storming protocol, and the bounded contexts in their final versions, purging the intermediate versions created during the refinement of the steps from the context.
A further improvement could be a more focused aggregate derivation, where the aggregates of each bounded context are derived in a separate \gls{llm} context, reducing the amount of context we need to supply to the specific bounded context. Together, we believe that these steps can considerably improve the accuracy of Step~4 in future work.

Step~5 shows multiple issues that need to be tackled in future work. First, generating PlantUML code directly for complex diagrams is not feasible. Automatic graph layout is a full research field in itself, and we cannot expect to generate good layouts with simple \gls{llm} prompts. However, layout can be seen as a downstream problem after generating the correct structure of the diagram.
Second, the experts complained that the technical architecture lacks sufficient detail and explanations for the system's implementation. We assume that this step is too complicated for today's generation of \glspl{llm}. There is a related work by Cervantes et al.~\cite{cervantes2025llmassistedapproachdesigningsoftware} that could fill this gap with another guided multi-prompt approach. They employ \glspl{llm} to assist with ADD, synthesizing a technical architecture. This will be subject to future work.

Overall, we can deduce a limited effectiveness in automating stages of the \gls{ddd} process, and reinforce the need for professional oversight. This answers our research question.

\subsection{Implications for Practice}
For the best results with this prompting framework, we recommend a detailed set of requirements and an expert guiding the \gls{llm}.

Without detailed requirements, the \glspl{llm} may struggle to provide meaningful results. The ubiquitous language may not fit the project well: The \gls{llm} may start to hallucinate or draw information from its internal world knowledge that might be inaccurate for the project, and without close expert guidance, these errors will propagate and multiply in the later steps.

Even with detailed requirements, the lack of expert guidance may compromise the results. Small errors generated in the first step can propagate and accumulate into larger errors in subsequent steps. Thus, we strongly advise judging and refining the generated artifacts rigorously.

Assuming these prerequisites, we recommend using this framework on one -- or multiple -- \glspl{llm} to generate a baseline for a discussion of the architecture. It helps direct the discussion to the important trade-offs and minimizes time spent on tedious tasks, such as creating a glossary for the ubiquitous language.

In many cases, it will be sufficient to follow Steps 1 to 3 rather than executing the full five-step framework, especially since the quality of the generated aggregates and technical architecture was subpar in our evaluation.

Summarizing the points raised in the evaluation and discussion, we recommend special care with the following points:
\begin{itemize}
    \item Bounded contexts might look reasonable, but may be too fine-grained in practice. They should be consolidated, where appropriate.
    \item The design of aggregates seems fragile in our evaluation. If the generated aggregates are inaccurate, fixing them might require more work than designing them from scratch.
    \item The technical architecture mapping was not particularly useful in our case. It might be advisable to omit the last step or to exercise it without \gls{llm} assistance.
\end{itemize}

\subsection{Limitations}

By the nature of the case study, the generalizability of the results is limited.

\paragraph{External validity}
Two points limit the external validity: First, this case study evaluates only three commercial \glspl{llm} in a specific point in time. Since the three models generated similar results, we can reasonably assume some generalizability to similar models. However, the results cannot be generalized to smaller \glspl{llm} or future \gls{llm}-generations.
Second, our case study was operating on detailed requirements written in German. Our results do not generalize to other projects with less detailed requirements. A generalization to requirements written in other languages that the \gls{llm} is proficient with can be reasonably assumed. Additionally, the limited context windows of \glspl{llm} hinder generalizability to requirements of much larger size.
\paragraph{Construct validity}
The case study evaluation was conducted by experts who are very familiar with the requirements. However, they also conducted \gls{ddd} on SecuRooms before and are familiar with the architecture of both SecuRooms and SecuMails. This can lead to a bias in their view of the generated design, especially since they tended to compare the generated artifacts to the existing SecuRooms architecture in the evaluation.
The FTAPI products are described on their website, so the \glspl{llm} could know about them from their training data. However, the technical requirements were not published at the time the prompting framework was exercised, so they could not have been part of the \glspl{llm}'s training data.

Another limitation concerns the concept of assisting with or automating \gls{ddd} itself: \Gls{ddd} was designed to foster collaboration among stakeholders. While automating parts of the process with our prompting framework could improve collaboration by making time spent more meaningful, it could also reduce collaboration in practice. This risk must be handled carefully by users of the prompting framework.

\subsection{Future Work}

In our future work, we want to address the shortcomings of the prompting framework. This includes the following facets:

We want to improve our prompts based on this evaluation, especially the prompts for Step~3 and Step~4 show some shortcomings that should be easily addressable. Step~5 was not as thoroughly evaluated, but we can assume that the prompt needs substantial work, either by replacing the step with a second process, such as ADD~\cite{cervantes2025llmassistedapproachdesigningsoftware}, by removing the step altogether, acknowledging that today's \glspl{llm} can not provide additional value for this step, or by refining the prompt in a way that successfully generates sophisticated technical architectures.

In this prompting framework, the consistency between the artifacts is not ensured. The \gls{llm} might use terms that are not defined in the glossary. Similarly, concepts that are discovered only in later steps do not propagate back to the glossary or the event storming. It would be beneficial to add some safeguards and refinement loops to this process.

While this case study shows the general, but limited, usefulness of the prompt framework, a larger evaluation is needed to confirm its usefulness before broader adoption becomes feasible. We aim to conduct such a larger evaluation with additional experts and more projects to improve the generalizability of our evaluation.

\section{Conclusion\generated{ (0.5 pages)}}\label{sec:conclusion}

Conducting steps of \gls{ddd} manually is a time-consuming task that requires a lot of expertise from software architects. To overcome this challenge, we propose a five-step prompting framework that can lead through the design process.
It can assist the architect with tedious tasks and offer different perspectives for consideration.
The comparative evaluation of the prompting framework with a real-world project designed using \gls{ddd} shows that the \glspl{llm} are capable of executing these steps.
However, the \glspl{llm} can struggle with complex requirements and more abstract tasks, especially designing aggregates and mapping technical architectures.
We assume that more automation potential can be uncovered with more sophisticated prompting and future \gls{llm}-generations.

Overall, the interviewed architecture experts agreed that the proposed \gls{ddd} prompting framework is useful in their work. They pointed out in which areas their design could have benefited from the \gls{llm}'s perspective, and in which areas the \gls{llm} started to hallucinate or produce unvaluable data.

\generated{
    Do:
        First paragraph: Summary of practitioner value (not academic contribution)
        Second paragraph: Clear next steps ("Start by implementing Phase 1 next week")
        End with: "The most important lesson: [actionable insight]"
    Critical: Remove all "future work" academic phrasing - replace with "What we're doing next month"
    Why it works: Satisfies CFP's requirement to "highlight... real-world applicability"
}

\generated{

\section{IEEE notes}

\subsection{Abbreviations and Acronyms}\label{AA}
Define abbreviations and acronyms the first time they are used in the text, 
even after they have been defined in the abstract. Abbreviations such as 
IEEE, SI, MKS, CGS, ac, dc, and rms do not have to be defined. Do not use 
abbreviations in the title or heads unless they are unavoidable.

\subsection{Some Common Mistakes}\label{SCM}
\begin{itemize}
\item The word ``data'' is plural, not singular.
\item Do not use the word ``essentially'' to mean ``approximately'' or ``effectively''.
\item In your paper title, if the words ``that uses'' can accurately replace the word ``using'', capitalize the ``u''; if not, keep using lower-cased.
\end{itemize}
}

\section*{Acknowledgment}

The authors used generative AI tools to assist with writing this paper:
Grammarly was used for spelling, grammar, and readability improvements.
The authors used meta-prompting to refine the proposed prompting framework.
The authors used the AI tool Turbo Scribe to create transcripts of the interviews.

All technical ideas, analyses, results, and conclusions in this paper were conceived, developed, and verified solely by the authors.
The authors take full responsibility for the content of the final manuscript.

\bibliographystyle{IEEEtran}
\bibliography{references}

\end{document}